\documentclass[preprint,aps,showpacs,preprintnumbers,amsmath,amssymb]{revtex4}  
\usepackage{graphicx}
\usepackage{dcolumn}
\usepackage{bm}

\begin{document}

\title{Multi-Channel Selective Femtosecond Coherent Control \\ Based on Symmetry Properties}

\author{Andrey Gandman, Lev Chuntonov, Leonid Rybak}
\author{Zohar Amitay}%
\email{amitayz@tx.technion.ac.il} %
\affiliation{Schulich Faculty of Chemistry, Technion - Israel Institute of Technology, Haifa 32000, Israel}

\begin{abstract}
We present and implement a new scheme for extended multi-channel
selective femtosecond coherent control based on symmetry properties
of the excitation channels. Here, an atomic non-resonant two-photon
absorption channel is coherently incorporated in a
resonance-mediated (2+1) three-photon absorption channel. By proper
pulse shaping, utilizing the invariance of the two-photon absorption
to specific phase transformations of the pulse, the three-photon
absorption is tuned independently over order-of-magnitude yield
range for any possible two-photon absorption yield. Noticeable is a
set of "two-photon dark pulses" inducing widely-tunable three-photon
absorption.
\end{abstract}


\maketitle

When a quantum system is irradiated with a broadband femtosecond pulse,
a coherent manifold of quantum pathways is photo-induced from one state to the other.
Shaping the pulse \cite{pulse_shaping} to manipulate 
the interferences among these pathways is
the means by which femtosecond coherent control affects 
state-to-state transition probabilities
\cite{tannor_kosloff_rice_coh_cont,shapiro_brumer_coh_cont_book,warren_rabitz_coh_cont,
rabitz_vivie_motzkus_kompa_coh_cont,dantus_exp_review1_2}.
When several 
excitation channels to different final states of the system are of concern,
in many cases it is sufficient to achieve basic multi-channel selective control ("on-off" type)
of maximizing one channel while minimizing the other channels.
However, in many other cases a much higher degree of selectivity is desirable, 
where ideally one channel is tuned independently over its full yield range  
while the other channels are set constant on any chosen possible yields. 
This requires pulse shapes that reduce the correlations between the different channel~yields.

Such extended multi-channel selective femtosecond control have been experimentally studied 
so far mostly by employing 
automatic "black-box" 
optimization of the pulse shape using learning algorithms
\cite{rabitz_feedback_learning_idea,gerber_feedback_control_review,levis_feedback_cont,motzkus_bio1_feedback_cont}.
Here we focus on rational femtosecond coherent control, where the pulse shaping is based on identifying first
the state-to-state interfering pathways and their interference mechanism.
By itself rational femtosecond control is suitable mainly to quantum systems of limited complexity,
however it might also serve as a basis for establishing 
control principles that will be incorporated in extended control schemes of high-complexity systems.
To date, the only rational femtosecond control work studied aspects of 
extended multi-channel selective control is the one by Dudovich {\it et al.}~\cite{silberberg-angular-dist}
on polarization control of two-photon absorption to 
different angular-momentum states.
All the other past studies of rational 
selective femtosecond control involve the simpler ("on-off") 
scheme \cite{gerber_Na_Na2,silberberg_antiStokes_Raman_spect,wollenhaupt-baumert1_2}. 

In this Letter we present and demonstrate a new rational scheme for
extended multi-channel selective femtosecond coherent control that
is based on symmetry properties of the excitation channels. The
processes under study here are multiphoton absorption processes that
are of fundamental and applicative importance 
and have been shown to be coherently controlled very effectively
\cite{dantus_exp_review1_2,silberberg-angular-dist,silberberg_antiStokes_Raman_spect,wollenhaupt-baumert1_2,
silberberg_2ph_nonres1_2,dantus_2ph_nonres_molec1_2,baumert_2ph_nonres,silberberg_2ph_1plus1,
girard_2ph_1plus1,becker_2ph_1plus1_theo,gersh_murnane_kapteyn_Raman_spect,
leone_res_nonres_raman_control,leone_cars,amitay_3ph_2plus1_2,amitay_2ph_inter_field1_2,silberberg-2ph-strong-field,
weinacht-2ph-strong-theo-exp}.
The present atomic scenario involves a non-resonant two-photon absorption channel
that is coherently incorporated in a resonance-mediated (2+1) three-photon absorption channel.
Their one-channel control 
have previously been studied separately \cite{silberberg_2ph_nonres1_2,amitay_3ph_2plus1_2}.
Here, by utilizing a symmetry property of the two-photon absorption for proper pulse shaping,
the three-photon absorption is tuned independently over order-of-magnitude yield range
for any possible yield of the two-photon absorption. 
The approach developed here is general, conceptually simple, and very effective. 

\begin{figure} [thbp]
\includegraphics[scale=2]{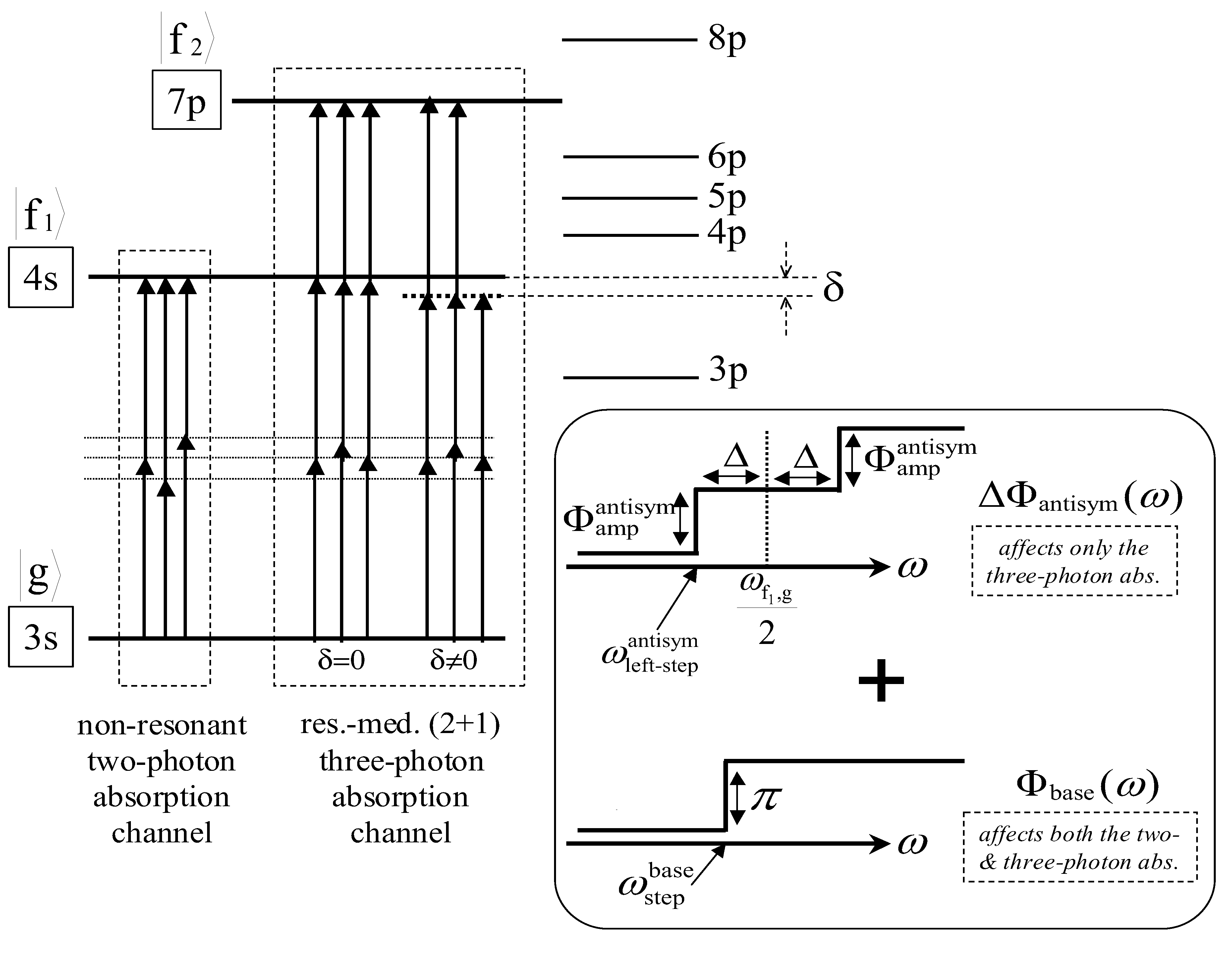}
\caption{\label{fig1}
The two-channel femtosecond excitation scheme of Na, including 
a non-resonant two-photon absorption channel from
$\left| g \right\rangle$$\equiv$$3s$ to $\left| f_1 \right\rangle$$\equiv$$4s$ and a
resonance-mediated (2+1) three-photon absorption channel
from $\left| g \right\rangle$$\equiv$$3s$ to $\left| f_2 \right\rangle$$\equiv$$7p$ via
$\left| f_1 \right\rangle$$\equiv$$4s$.
Shown are examples of 
two- and three-photon pathways.
The latter 
are either on resonance or near resonance with $\left| f_1 \right\rangle$ 
(with detuning $\delta$).
The inset schematically shows the phase patterns used in the selective coherent control scheme.
The base patterns $\Phi_{\textrm{base}}(\omega)$ are of a single $\pi$ step at
variable position $\omega_{\textrm{step}}^{\textrm{base}}$.
The anti-symmetric additions $\Delta\Phi_{\textrm{antisym}}(\omega)$
are composed of two steps of variable amplitude $\Phi_{\textrm{amp}}^{\textrm{antisym}}$
positioned symmetrically around $\omega_{f_1,g}/2$ at variable positions that are represented 
by the left-step position $\omega_{\textrm{left-step}}^{\textrm{antisym}}$.
} \end{figure}

The two-channel excitation considered here is shown schematically in Fig.~\ref{fig1}.
It involves an initial ground state $\left|g\right>$ and
two excited states $\left|f_{1}\right>$ and $\left|f_{2}\right>$.
The $\left|g\right>$ and $\left|f_{1}\right>$ states are coupled by
a non-resonant two-photon coupling provided by a manifold of states
$\left|v\right>$ 
that are far from resonance.
The $\left|f_{1}\right\rangle$ and $\left|f_{2}\right\rangle$ states
are coupled resonantly by one-photon coupling.
Hence, when irradiated with a weak (shaped) femtosecond pulse 
a non-resonant two-photon absorption 
from $\left|g\right>$ to $\left|f_{1}\right>$ is induced simultaneously with
a resonance-mediated (2+1) three-photon absorption 
from $\left|g\right>$ to $\left|f_{2}\right>$ with $\left|f_{1}\right>$ as an intermediate state.
The final (complex) amplitudes $A_{f_1}$ and $A_{f_2}$ of states $\left|f_{1}\right>$ and $\left|f_{2}\right>$
after the pulse is over are given, respectively, by 2$^{nd}$- and 3$^{rd}$-order time-dependent
perturbation theory as  \cite{silberberg_2ph_nonres1_2,amitay_3ph_2plus1_2}
\begin{equation}
A_{f_1} = -\frac{1}{i \hbar^2} \mu_{f_1,g}^2 A^{(2)}(\omega_{f_1,g}) ,
\label{eq_Af1}
\end{equation}
and
\begin{eqnarray}
A_{f_2} & = & \frac{1}{\hbar^3} \mu_{_{f_2,f_1}} \mu_{f_1,g}^2 \left[ A_{f_2}^{(2+1)on-res} + A_{f_2}^{(2+1)near-res} \right] ,
\label{eq_Af2}
\\
A_{f_2}^{(2+1)on-res} & = & 
i\pi E(\omega_{f_2,f_1})A^{(2)}(\omega_{f_1,g}) \; ,
\label{eq_Af2_on_res}
\\
A_{f_2}^{(2+1)near-res} & = &
-\wp\int_{-\infty}^{\infty} \frac{1}{\delta} A^{(2)}(\omega_{f_1,g}-\delta) E(\omega_{f_2,f_1}+\delta) d\delta \; ,
\label{eq_Af2_near_res}
\end{eqnarray}
with
\begin{eqnarray}
A^{(2)}(\Omega) & = & \int_{-\infty}^{\infty} E(\omega) E(\Omega-\omega)d\omega
= \int_{-\infty}^{\infty} E(\Omega/2-\alpha) E(\Omega/2+\alpha)d\alpha
\; . \label{eq_A2}
\end{eqnarray}
The quantities $\mu_{f_2,f_1}$ and $\mu_{f_1,g}^2$ are, respectively,
the $\left|f_2\right\rangle$-$\left|f_1\right\rangle$ one-photon dipole matrix element
and $\left|f_{1}\right\rangle$-$\left|g\right\rangle$ effective non-resonant
two-photon dipole coupling \cite{silberberg_2ph_nonres1_2},
and $\omega_{m,n}$ is the transition frequency between a pair of states. 
The 
$E(\omega) \equiv \left|E(\omega)\right| \exp \left[ i\Phi(\omega) \right]$ is the pulse spectral field,
related to the temporal field by Fourier transform,
with $\left|E(\omega)\right|$ and $\Phi(\omega)$ being, respectively, the spectral amplitude and phase of frequency $\omega$.
For the (unshaped) transform-limited pulse,
which is the shortest pulse for a given spectrum $\left|E(\omega)\right|$,
$\Phi(\omega)=0$ for any $\omega$.
The final populations 
of $\left|f_1\right\rangle$ and $\left|f_2\right\rangle$
are given, respectively, by $P_{f_1} = \left|A_{f_1}\right|^{2}$ and $P_{f_2} = \left|A_{f_2}\right|^{2}$.
They serve as the measures for the two-photon and three-photon absorption.

The final amplitude $A_{f_1}$ of $\left|f_1\right\rangle$ coherently
interferes all the possible non-resonant two-photon pathways from $\left|g\right\rangle$ to $\left|f_1\right\rangle$,
i.e., coherently integrates over all their corresponding amplitudes.
Each such pathway is composed of two absorbed photons of frequencies $\omega$ and $\omega_{f_1,g}-\omega$.
The final amplitude $A_{f_2}$ of $\left|f_2\right\rangle$ coherently
interferes all the possible resonance-mediated (2+1) three-photon pathways
from $\left|g\right\rangle$ to $\left|f_2\right\rangle$.
Each such pathway is either on resonance or near resonance with 
$\left|f_1\right\rangle$, having a corresponding detuning $\delta$.
It involves a non-resonant absorption of two photons with a
two-photon transition frequency $\omega_{f_1,g}-\delta$ and the
absorption of a third complementary photon of frequency
$\omega_{f_2,f_1}+\delta$.
The term $A_{f_2}^{(2+1)on-res}$ interferes all the on-resonant pathways ($\delta=0$),
while the term $A_{f_2}^{(2+1)near-res}$ interferes all the near-resonant pathways ($\delta\ne0$)
with a $1/\delta$ amplitude weighting. 
The on-resonant pathways are excluded from $A_{f_2}^{(2+1)near-res}$ by
the Cauchy's principal value operator $\wp$.
Several two- and three-photon pathways are shown schematically in Fig.~\ref{fig1}.
The different amplitudes are expressed using the parameterized
amplitude $A^{(2)}(\Omega)$ interfering all the possible two-photon
pathways with transition frequency $\Omega$,
each 
composed of two photons with frequencies $\omega$ and $\Omega-\omega$, or, equivalently,
$\Omega/2-\alpha$ and $\Omega/2+\alpha$, i.e., two frequencies located symmetrically around $\Omega/2$.
The amplitudes $A_{f_1}$ and $A_{f_2}^{(2+1)on-res}$ are proportional to
$A^{(2)}(\Omega=\omega_{f_1,g})$, 
while $A_{f_2}^{(2+1)near-res}$ integrates over
all the $A^{(2)}(\Omega=\omega_{f_1,g}-\delta)$ 
with non-zero detuning ($\delta \ne 0$).

In order to achieve high degree of selectivity between the two- and three-photon absorption channels we
utilize phase transformations $\hat{U}^{(\textrm{selective-cntrl})}_{\textrm{phase}}$ of the 
spectral field $E(\omega)$ that do not change
the two-photon absorption amplitude $A_{f_1}$ but do change the three-photon absorption amplitude $A_{f_2}$.
Based on Eqs.~(\ref{eq_Af1})-(\ref{eq_A2}), 
it applies to any transformation
\begin{equation}
\hat{U}^{(\textrm{selective-cntrl})}_{\textrm{phase}} = \exp[i \Delta\Phi_{\textrm{antisym}}(\omega)]
\end{equation}
that corresponds 
to the addition of a spectral phase pattern
$\Delta\Phi_{\textrm{antisym}}(\omega)$$\equiv$$\Delta\Phi_{\textrm{antisym}}(\omega_{f_1,g}/2+\alpha)$
that is anti-symmetric around $\omega_{f_1,g}/2$, i.e., for any value of $\alpha$ it satisfies the relation
\begin{equation}
\Delta\Phi_{\textrm{antisym}}(\omega_{f_1,g}/2+\alpha) = -\Delta\Phi_{\textrm{antisym}}(\omega_{f_1,g}/2-\alpha) \; .
\end{equation}
Such phase addition keeps the value of $A^{(2)}(\Omega=\omega_{f_1,g})$ unchanged,
while it generally changes the value of $A^{(2)}(\Omega=\omega_{f_1,g}-\delta)$ for $\delta$$\ne$0.
Hence, it alters $A_{f_2}$ while keeping $A_{f_1}$ invariant.

So, the following extended selective coherent control scheme 
for designing the spectral phase pattern $\Phi(\omega)$  
can be applied: The two-photon absorption amplitude $A_{f_1}$ is set to a chosen value $A_{f_1,\textrm{base}}$
by choosing a proper base phase pattern $\Phi_{\textrm{base}}(\omega)$.
Then, by adding different suitable anti-symmetric phase patterns
$\Delta\Phi_{\textrm{antisym}}(\omega)$, the three-photon absorption
amplitude $A_{f_2}$ is tuned from its base value
$A_{f_2,\textrm{base}}$ over a wide range of values, while $A_{f_1}$
is kept constant on $A_{f_1,\textrm{base}}$.
The total spectral phase pattern applied to the shaped pulse is
$\Phi(\omega) = \Phi_{\textrm{base}}(\omega)+
\Delta\Phi_{\textrm{antisym}}(\omega)$.

The phase patterns chosen here as $\Phi_{\textrm{base}}(\omega)$ and $\Delta\Phi_{\textrm{antisym}}(\omega)$
are shown schematically in Fig.~\ref{fig1}.
The present base patterns $\Phi_{\textrm{base}}(\omega)$ are of a single $\pi$ step
at variable position 
$\omega_{\textrm{step}}^{\textrm{base}}$.
As previously shown \cite{silberberg_2ph_nonres1_2}, this family of pulse shapes
allows high degree of the
control over the full range of the
non-resonant two-photon absorption channel.
It ranges from zero absorption, corresponding to 
"two-photon dark pulses",
up to the maximal possible absorption, corresponding to the TL pulse
that induces fully constructive interferences among all the 
two-photon pathways [see Eqs.(\ref{eq_Af1}) and (\ref{eq_A2})].
In other separate studies \cite{amitay_3ph_2plus1_2}, the $\pi$-step patterns have also been shown to be
very effective in controlling the resonance-mediated (2+1) three-photon absorption channel.
The present anti-symmetric phase additions $\Delta\Phi_{\textrm{antisym}}(\omega)$ are composed of
two steps of variable amplitude $\Phi_{\textrm{amp}}^{\textrm{antisym}}$
positioned symmetrically around $\omega_{f_1,g}/2$ at variable positions that are represented 
by the left-step position $\omega_{\textrm{left-step}}^{\textrm{antisym}}$.
%


The physical model system of the study is the sodium (Na) atom \cite{NIST},
with the $3s$ ground state as $\left|g\right\rangle$, the
$4s$ state as $\left|f_1\right\rangle$, and the $7p$ state as
$\left|f_2\right\rangle$ (see Fig.~\ref{fig1}).
The transition frequency $\omega_{f_1,g} \equiv \omega_{4s,3s} = 25740$~cm$^{-1}$ corresponds to two 777-nm photons and the
transition frequency $\omega_{f_2,f_1} \equiv \omega_{7p,4s} = 12801$~cm$^{-1}$ corresponds to a 781.2-nm photon.
The $3s$-$4s$ non-resonant two-photon coupling originates from the manifold of $p$-states,
particularly from the $3p$ state [$\omega_{3p,3s}$$\sim$16978~cm$^{-1}$ (589~nm)].
The sodium is irradiated with phase-shaped linearly-polarized femtosecond pulses having a Gaussian
intensity spectrum centered around 780~nm (12821~cm$^{-1}$) with 5.8-nm (95-cm$^{-1}$) bandwidth
($\sim$180-fs TL duration).

Experimentally, a sodium vapor in a heated cell is irradiated with such laser pulses, after
they undergo shaping in an optical setup incorporating a pixelated
liquid-crystal spatial light phase modulator \cite{pulse_shaping}.
The effective spectral
shaping resolution is $\delta\omega_{shaping}$=2.05~cm$^{-1}$ 
per pixel.
The peak intensity of the TL pulse is below 10$^{9}$~W/cm$^{2}$. 
Following the interaction with a pulse, the population excited to the $4s$ state radiatively decays to the
lower $3p$ state, which then decays to the $3s$ ground state.
The $3p$-$3s$ fluorescence serves as the relative measure for the total $4s$ population $P_{f_1} \equiv P_{4s}$.
The population excited to~the $7p$ state undergoes radiative and collisional decay to lower excited states,
including the $4d$, $5d$, $6d$, and $6s$ states.
The fluorescence emitted in their decay to the $3p$ state 
serves as the relative measure for the total $7p$ population $P_{f_2} \equiv P_{7p}$.
The fluorescence is measured 
using a spectrometer coupled to a time-gated camera system.
The $3p$-$3s$ fluorescence part originating from the $4s$ state is discriminated from the part originating from the $7p$ state
by using a proper 
detection gate width, utilizing
the different~time scales of the $4s$-to-$3p$ and $7p$-to-$3p$ decays.

\begin{figure} [thbp]
\includegraphics[scale=0.7]{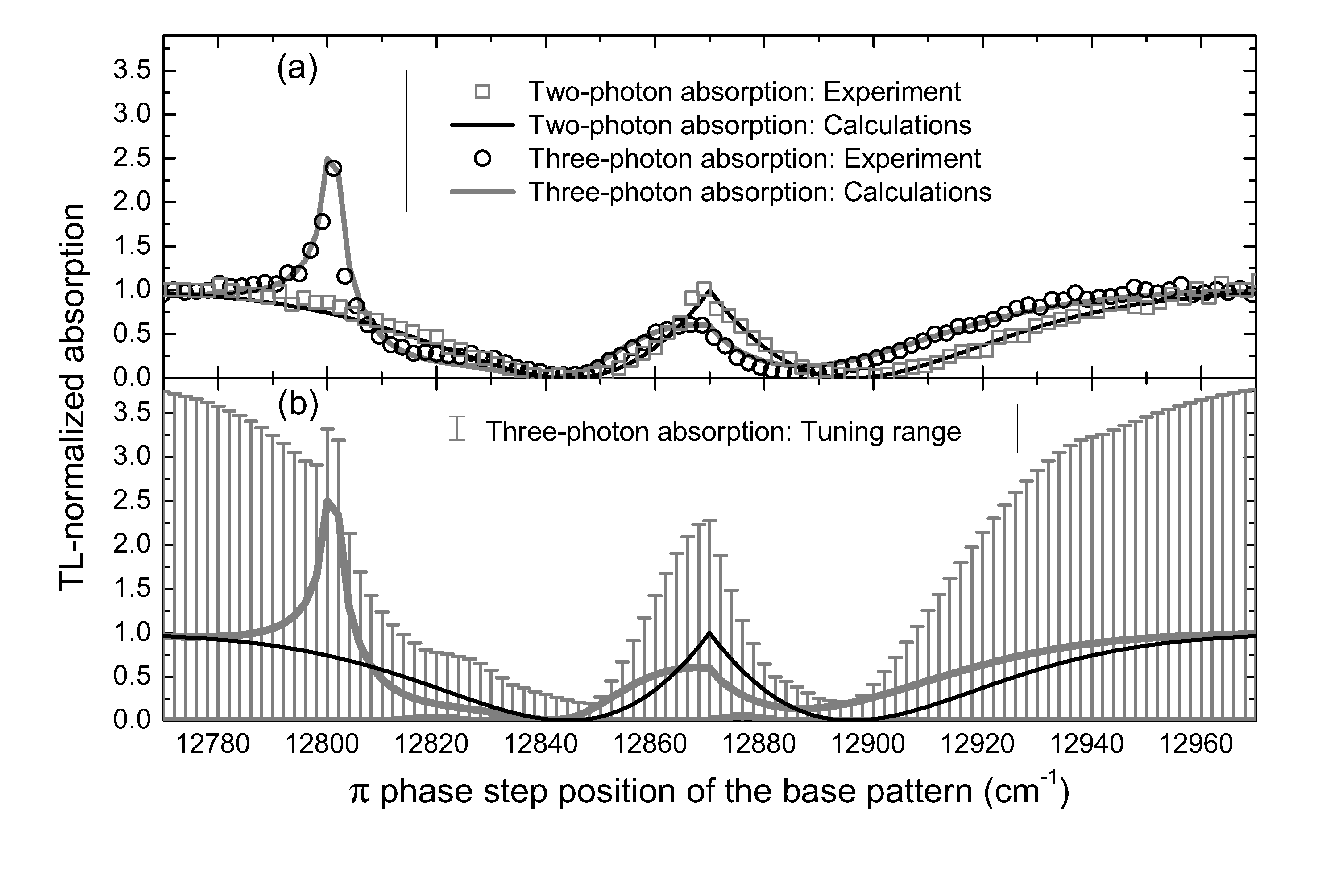}
\caption{\label{fig2}
(a) Experimental (circles and squares) and numerical-theoretical (lines) results for the
TL-normalized non-resonant two-photon absorption (squares and black line)
and resonance-mediated (2+1) three-photon absorption (circles and gray line) in Na,
as a function of the $\pi$-step position $\omega_{\textrm{step}}^{\textrm{base}}$
of the base patterns $\Phi_{\textrm{base}}(\omega)$.
(b) The complete picture of the extended symmetry-based selective control of the two-channel absorption in Na.
The lines are the numerical-theoretical results for the two- and three-photon absorption set by the base phase patterns,
presented also in panel (a).
The bar around each three-photon absorption point indicates the 
extended tuning range of the three-photon absorption
achieved by applying all the possible double-step anti-symmetric phase additions,
while the two-photon absorption is kept constant on its corresponding base level. 
} \end{figure}

Figure~\ref{fig2}(a) presents experimental (circles and squares) and theoretical (lines) results
for the basic phase control of the two-channel absorption in Na using the set of shaped pulses
with the base phase patterns $\Phi_{\textrm{base}}(\omega)$ of a single $\pi$ step.
Shown are $P_{4s}$ for the non-resonant two-photon absorption (squares and black line)
and $P_{7p}$ for the resonance-mediated (2+1) three-photon absorption (circles and gray line)
as a function of the $\pi$-step position $\omega_{\textrm{step}}^{\textrm{base}}$.
Each trace is normalized by the corresponding $P_{4s}$ or
$P_{7p}$~excited~by~the~TL~pulse.~The theoretical results are
calculated numerically using Eqs.(\ref{eq_Af1})-(\ref{eq_A2}), using
a grid with a bin size equal to the experimental shaping resolution
$\delta\omega_{shaping}$.
As can be seen, there is an excellent 
agreement between the experimental results and
the numerical-theoretical results. 
The non-resonant two-photon absorption is controlled from zero to 100\% of the absorption
induced by the TL pulse \cite{silberberg_2ph_nonres1_2}.
The zero absorption corresponds to the two dark pulses occurring when $\omega_{\textrm{step}}^{\textrm{base}}$=12896  
and 12844~cm$^{-1}$. 
Its upper limit is measured here to be~300-fold~smaller~than the TL absorption,  
as set by the best-achieved experimental noise.
%
The resonance-mediated (2+1) three-photon absorption is experimentally controlled from 3\% to 240\% of the 
TL absorption \cite{amitay_3ph_2plus1_2}.
The strong enhancement occurs when $\omega_{\textrm{step}}^{\textrm{base}}$=$\omega_{7p,4s}$=12801~cm$^{-1}$.
As previously identified \cite{amitay_3ph_2plus1_2},
it originates from a change in the nature of the interferences
between the positively-detuned ($\delta$$>$0) and negatively-detuned ($\delta$$<$0) near-resonant $3s$-$7p$
three-photon pathways. 
With the TL pulse they are destructive, while with a $\pi$-step at $\omega_{7p,4s}$ they are constructive.
Overall, as seen from the results, the $\pi$-step position simultaneously sets the
two-photon and three-photon absorption levels to specific correlated values.

\begin{figure} [htbp]
\includegraphics[scale=0.9]{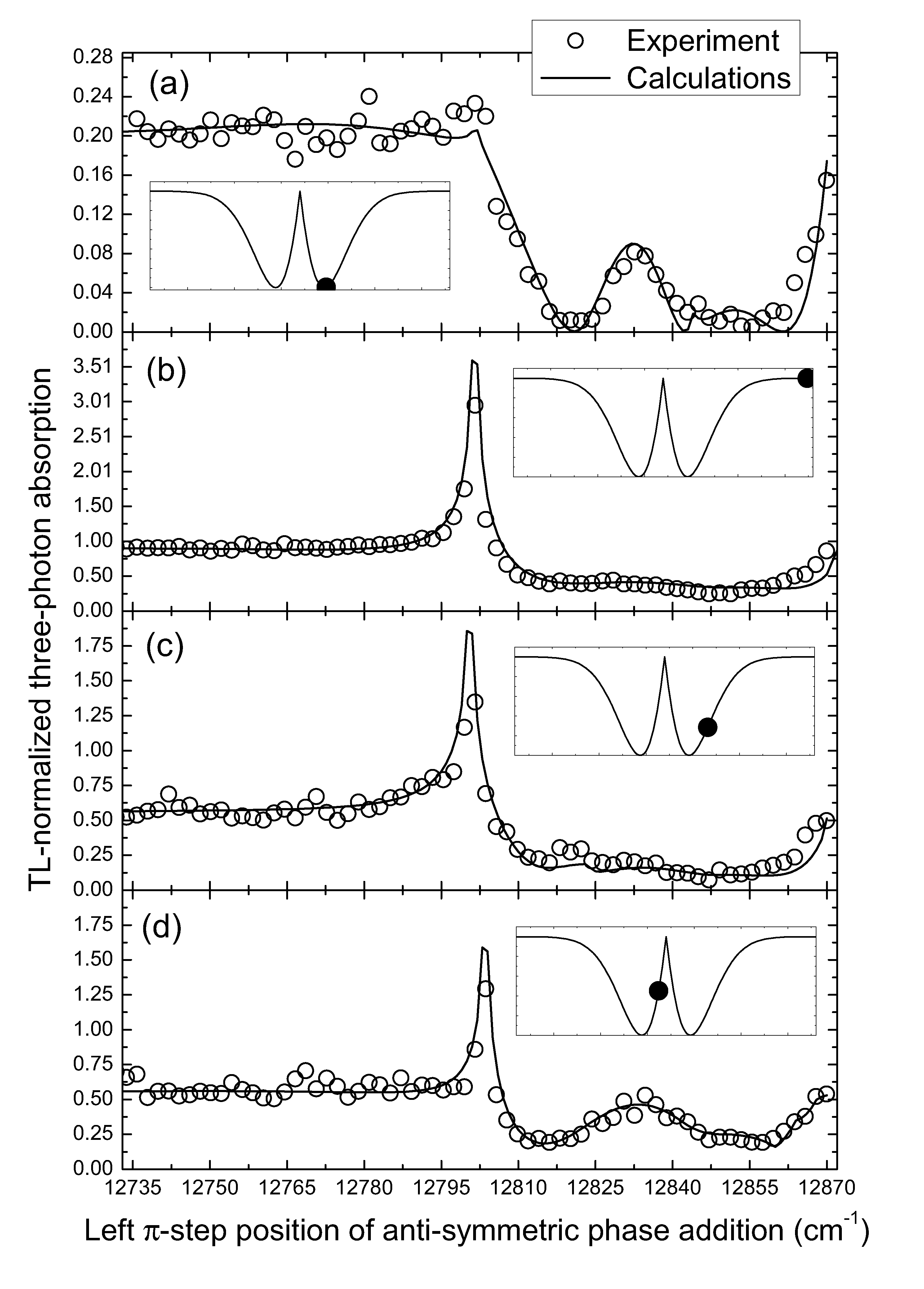}
\caption{\label{fig3} Experimental (circles) and
numerical-theoretical (lines) results for several example cases of
the symmetry-based selective coherent control implementation. Each
panel corresponds to a different position of the base $\pi$-step
$\omega_{\textrm{step}}^{\textrm{base}}$ of
$\Phi_{\textrm{base}}(\omega)$, setting the non-resonant two-photon
absorption to a different chosen level indicated in the panel's
inset.
The main graph in each panel displays the TL-normalized resonance-mediated (2+1) three-photon absorption
resulting from different anti-symmetric phase additions $\Delta\Phi_{\textrm{antisym}}(\omega)$
with a double $\pi$-step ($\Phi_{\textrm{amp}}^{\textrm{antisym}}$=$\pi$)
at different positions represented by the left $\pi$-step position $\omega_{\textrm{left-step},\pi}^{\textrm{antisym}}$.
The panels correspond to:
(a) zero two-photon absorption, i.e., to a family of "two-photon dark pulses";
(b) maximal two-photon absorption, i.e., the TL absorption;
and (c) and (d) intermediate levels of the two-photon absorption.
} \vspace{-5cm}
\end{figure}

The implementation of the 
symmetry-based selective coherent control scheme presented above
strongly reduces this correlation. Several examples are shown in Fig.~\ref{fig3}.
Each panel corresponds to a different position of the base $\pi$-step $\omega_{\textrm{step}}^{\textrm{base}}$ of
$\Phi_{\textrm{base}}(\omega)$, setting the non-resonant two-photon absorption 
to a different chosen level ($P_{4s,\textrm{base}}$),
as indicated in the panel's inset.
The main graph in each panel displays the TL-normalized resonance-mediated (2+1) three-photon absorption ($P_{7p}$)
resulting from 
different double-step anti-symmetric phase additions
$\Delta\Phi_{\textrm{antisym}}(\omega)$ with
$\Phi_{\textrm{amp}}^{\textrm{antisym}}$=$\pi$. It is presented as a
function of the left $\pi$-step position
$\omega_{\textrm{left-step},\pi}^{\textrm{antisym}}$. Shown are
experimental (circles) and numerical-theoretical (lines)~results.

Figure~\ref{fig3}(a) shows the results for the case of zero
two-photon absorption set by
$\omega_{\textrm{step}}^{\textrm{base}}$=12896~cm$^{-1}$.
By scanning $\omega_{\textrm{left-step},\pi}^{\textrm{antisym}}$ across the spectrum,
the three-photon absorption is continuously tuned experimentally
from below 1\% up to about 20\% of the TL absorption,
while the two-photon absorption is kept constant on its zero level
that experimentally corresponds to a noise-level signal (see above).
In other words, the results of Fig.~\ref{fig3}(a) correspond to a family of shaped pulses that are all
dark with respect to the two-photon~absorption, while each of them induces a different 
three-photon absorption level tunable over an order-of-magnitude range of values.
The present tuning range 
is determined only by~the near-resonant
component $A_{7p}^{(2+1)near-res}$ of the three-photon absorption amplitude, since a two-photon dark pulse also 
leads to a zero on-resonant component $A_{7p}^{(2+1)on-res}$~[see~Eqs.(\ref{eq_Af1})-(\ref{eq_A2})].

Figure~\ref{fig3}(b) shows the results for the case of maximal two-photon absorption, i.e., the TL absorption.
It is set by $\omega_{\textrm{step}}^{\textrm{base}}$ that is outside the spectrum, leading to a base phase
pattern that is constant across the whole spectrum.
By changing the $\pi$-step positions of the double-step anti-symmetric phase addition,
the three-photon absorption is continuously tuned experimentally from $\sim$30\% up to $\sim$300\%
of the TL absorption, while the two-photon absorption is kept constant on its maximal level.
The strong enhancement of the resonance-mediated (2+1) three-photon absorption occurs when
$\omega_{\textrm{left-step},\pi}^{\textrm{antisym}} = \omega_{7p,4s}$.
The corresponding interference mechanism is similar to the one described above for the enhancement
occurring with a single $\pi$-step at $\omega_{\textrm{step}}^{\textrm{base}} = \omega_{7p,4s}$: 
the introduction of such double anti-symmetric $\pi$-step leads to a
destructive-to-constructive change in the nature of the interferences among near-resonant $3s$-$7p$ three-photon pathways.
The higher enhancement induced with the anti-symmetric double-step (300\% vs.~240\% of the TL absorption)
is due to the fact that the anti-symmetric phase addition keeps
the on-resonant component $A_{7p}^{(2+1)on-res}$ on its maximal magnitude set by the uniform base pattern,
while a single $\pi$-step at $\omega_{\textrm{step}}^{\textrm{base}} = \omega_{7p,4s}$ sets it to a
lower magnitude \cite{amitay_3ph_2plus1_2}.
Figures~\ref{fig3}(c)-(d) present additional selective control results where the two-photon absorption is set on
constant intermediate levels in-between zero and the TL absorption, while the three-photon absorption is
independently tuned experimentally from $\sim$10\% [Fig.~\ref{fig3}(c)] or $\sim$20\% [Fig.~\ref{fig3}(d)]
up to $\sim$135\% of the TL absorption.

The complete picture for the extended selective femtosecond control of the two-channel absorption
in Na is shown in Fig.~\ref{fig2}(b), presenting numerical-theoretical results for the TL-normalized
non-resonant two-photon absorption ($P_{4s}$) and resonance-mediated (2+1) three-photon absorption ($P_{7p}$).
The two- and three-photon absorption levels set by a base $\pi$-step at a given $\omega_{\textrm{step}}^{\textrm{base}}$
are shown, respectively, by black and gray lines.
These are the same theoretical results shown in Fig.~\ref{fig2}(a).
The bar around each three-photon absorption point indicates the 
extended tuning range of the three-photon absorption
achieved by applying all the possible double-step anti-symmetric phase additions 
of any 0-2$\pi$ step amplitude $\Phi_{\textrm{amp}}^{\textrm{antisym}}$ 
and any steps' position (i.e., any $\omega_{\textrm{left-step}}^{\textrm{antisym}}$),
while the two-photon absorption is kept constant on its corresponding base level 
(black line).
As can be seen from this complete picture, for each two-photon absorption yield set by the
base phase pattern, the three-photon absorption yield can independently be tuned over a range of values
spanning one-two orders of magnitude.
Hence, high degree of independent selective control is achieved among the two absorption channels
and their degree of correlation is considerably reduced as compared to using only the base $\pi$-step patterns.
Quantitatively,  
the base levels of the two- and three-photon absorption as well as the detuning range
of the three-photon absorption depend on the spectral bandwidth of the pulse \cite{amitay_3ph_2plus1_2}.
Still, qualitatively, comparable degree of high selectivity among the two channels
is achieved with any spectral bandwidth. 


In conclusion, we have introduced and demonstrated a new scheme for achieving high-degree
of selective femtosecond coherent control among multiple excitation channels.
The approach developed here is general, conceptually simple, and very effective.
By proper pulse design, exploiting a symmetry property of one or more channels,
the channels of symmetry are set constant on chosen yields while the other
channels are tuned independently over a wide range of yields.
When many channels are involved, the scheme can be applied iteratively to utilize different
symmetry properties of different channels.
Amplitude and polarization shaping \cite{pulse_shaping} are natural extensions to the
phase shaping employed here.
Once corresponding symmetry properties are identified,
the new scheme can be used in various multi-channel scenarios
with various types of processes,
including also photo-excitations that lead to ionization and/or dissociation.

This research was supported by The Israel Science Foundation (grant No.~127/02),
by The James Franck Program in Laser Matter Interaction,
and by The Technion's Fund for The Promotion of Research.



\newpage

\end{document}